# Prigogine-Defay ratio of glassy freezing scales with liquid fragility

A. Loidl,[1,*] P. Lunkenheimer,[1] and K. Samwer[2]

[1] Experimental Physics V, University of Augsburg, 86135 Augsburg, Germany

[2] I. Physikalisches Institut, University of Göttingen, Göttingen, Germany

**A detailed study of published experimental data for a variety of materials on the incremental variation of heat capacity, thermal expansion, and compressibility at glassy freezing reveals a striking dependence of the Prigogine-Defay ratio *R* on the fragility index *m*. At high *m*, *R* approaches values of ~1, the Ehrenfest expectation for 2nd order continuous phase transitions, while *R* reaches values > 20 for low fragilities. We explain this correlation by the degree of separation of the glassy freezing temperature from a hidden phase transition into an ideal low-temperature glass.**

     The glass transition as function of temperature or pressure is a unique phenomenon of continuous solidification of a supercooled liquid avoiding crystallization. It occurs in a large variety of materials of high technological importance, ranging from silicate glasses, amorphous metals, and ionic salts to polymers. Glasses combine the mechanical stability of crystals with the disorder of liquids. Despite considerable experimental and theoretical progress, the microscopic understanding of the glass transition remains an unsolved and controversially discussed problem. The canonical glass-transition temperature $T_g$ is not a critical temperature in a thermodynamic sense, but marks a kinetic arrest, a smeared-out freezing-in of structural dynamics, below which amorphous materials are too viscous to flow on experimentally accessible time scales. Introductory reviews on supercooled liquids and glasses can be found in Refs. [1,2,3,4,5,6,7,8], to name but a few. Key features characterizing a supercooled liquid approaching $T_g$ are the massive increase of viscosity or structural relaxation time, the strong temperature dependence of entropy, which finally leads to an entropy crisis [9], and the development of dynamic heterogeneity characterized by a distribution of relaxation times [10]. The complex and remarkable features of glassy freezing were explained by different concepts, which can be grouped [1] into theories assuming an underlying phase transition into an "ideal" equilibrium glass state, either at $T_c < T_g$ [11,12,13,14,15] or at $T_c > T_g$ [16,17], and models not considering such hidden transitions [18,19,20,21].

     In the present work, we adapt the view of entropy models [11,12,13] assuming a vanishing configurational entropy below $T_c$. They belong to the first group, postulating a phase transition below $T_g$ underlying glass formation, which cannot be reached because of the freezing-in of structural dynamics at $T_g$. In recent years, such entropy models have gained considerable importance. Ideas of a critical temperature responsible for glass formation already started with Kauzmann's proposal [9] of an entropy catastrophe, where an extrapolation of the experimentally determined entropy of the glass would fall below the entropy of the crystal at a temperature, which is nowadays called Kauzmann temperature $T_K$. However, critical dynamical phenomena

accompanying the proposed structural phase transition cannot be accessed experimentally, but depend on extrapolations of quantities determined in the supercooled-liquid regime. Notably, critical remarks and questions concerning the validity of the Kauzmann paradox were raised in [22,23].

An adequate way to extrapolate structural dynamics of the supercooled liquid into the solid glass is the Vogel-Fulcher-Tammann (VFT) law [24,25,26,27] describing (at least approximately) the super-Arrhenius behavior of relaxation times $\tau$ (or viscosity $\eta \propto \tau$) in a broad temperature regime [1,28,29]:

$$\tau = \tau_0 \, exp\left[D\,T_{VF}/(T - T_{VF})\right] \qquad (1)$$

Here $\tau_0$ represents a microscopic time corresponding to an inverse attempt excitation frequency and $D$ is the strength parameter quantifying the degree of deviations from an Arrhenius behavior [27]. $T_{VF}$ is the Vogel-Fulcher temperature, where the relaxation times or viscosities are predicted to diverge. Notably, for a large variety of materials, the ratio $T_K/T_{VF}$ was found to be close to unity [5,30,31]. Hence, in the following we always will use $T_{VF}$ to estimate the critical temperature of the ideal glass transition. The VFT law can be rationalized using the Adam-Gibbs theory [12] assuming that supercooled liquids decay into cooperatively rearranging sub-regions, which relax independently and whose sizes diverge at the critical temperature. These ideas were further developed in a mean-field variant of an ideal structural glass transition, the random first-order phase transition [13]. Indeed, in a series of experimental and theoretical works it has been proven that the super-Arrhenius behavior of many glass-forming liquids close to $T_g$ is due to diverging length scales, which can be directly measured via higher-order susceptibilities [32,33,34,35,36]. One should note that numerous alternatives for the VFT law were proposed, but, in our view, it represents a convenient and common way to estimate the divergence temperature predicted by various theories [28,29]. A more detailed discussion is given in the Supplemental Material ([37], see also references therein [38,39,40,41,42,43,44,45,46,47,48,49,50,51,52,53,54,55,56,57,58, 59,60,61,62,63,64,65,66]).

The glass-transition temperature, as canonically defined, is characterized by an anomaly in the temperature dependence of specific heat, when measured with a cooling rate of 10 – 20 K/min or, alternatively, by the temperature where the mean relaxation time reaches a value of ~100 s, or the viscosity ~$10^{13}$ Poise. The glass transition is a kinetic phenomenon and, hence, $T_g$ depends on the cooling rate. Moreover, it only is visible by a change in the temperature dependence of volume $V$ or entropy $S$, without any latent heat contributions. This is schematically indicated in Fig. S1 [37]. At the coexistence line (see inset of Fig. S1 [37]), entropy and volume in the two phases are identical, but discontinuities appear in the second derivatives, like heat capacity, thermal expansion, or compressibility. In this respect, glassy freezing resembles a 2$^{nd}$ order phase transition, and one may ask whether the Ehrenfest relations derived for such transitions are applicable. Ehrenfest has shown that, for the inverse pressure dependence of the critical temperature $T_c$, the following two equations must be valid [67]:

$$dp/dT_c = \Delta c_p / T_c V_c \Delta \alpha_p \equiv A \qquad (2)$$

$$dp/dT_c = \Delta \alpha_p / \Delta \kappa_T \equiv B \qquad (3)$$

Here $V_c$ is the specific volume at $T_c$, while $\Delta c_p$, $\Delta \alpha_p$, and $\Delta \kappa_T$ are the discontinuities in heat capacity, thermal expansion, and compressibility, respectively, when crossing the phase boundary. $A$ and $B$ denote the right-hand terms in these equations, which will be discussed later. Equation (2) is derived from an isobaric path across the phase transition, assuming the continuity of entropy at $T_c$, while Eq. (3) follows from an isothermal path, assuming continuity of the volume. Prigogine and Defay [68] concluded that similar relations, leading to

$$R = \frac{\Delta c_p \, \Delta \kappa_T}{V_g \, T_g \, (\Delta \alpha_p)^2} = 1, \qquad (4)$$

(assuming $A = B$) should also apply to the glass transition, where $V_g$, $\Delta c_p$, $\Delta \alpha_p$, and $\Delta \kappa_T$ are the corresponding quantities at $T_g$. $R$ nowadays is called the Prigogine-Defay ratio (PDR).

The analysis of existing experimental data on the validity of Eq. (4) revealed that Eq. (2) (with $T_c = T_g$ and $V_c = V_g$) is correct and holds for most of the glass-forming materials studied, while Eq. (3) often is invalid, resulting in PDRs $R \geq 1$ [69,70,71,72,73,74,75]. The violation of Eq. (4) in glasses was explained in the framework of an order-parameter theory assuming that more than one order parameter has to be taken into account [69,70,71,72,76]. Several critical remarks on this interpretation and alternative explanations were proposed [77,78,79,80,81,82], specifically stating that glassy freezing is a smeared-out kinetic process and not a well-defined thermodynamic transition.

In the framework of the strong-fragile classification scheme of glasses as introduced by Angell [27], nowadays the steepness or fragility index $m$ is commonly used to quantify the deviations of $\tau(T)$ [or $\eta(T)$] from Arrhenius behavior. It is defined [83] by the slope at $T_g$ of the temperature-dependent relaxation times in an Angell plot [84], a $T_g$-scaled Arrhenius plot, $\log_{10}(\tau)$ vs. $T_g/T$. $m$ is related to the strength parameter in Eq. (1) via $m = 16 + 590/D$ [83]. Interestingly, this quantity can also be regarded as a measure how close the glass-transition temperature approaches $T_{VF}$. This is visualized in Fig. 1(a) showing an Angell plot extending to temperatures far below the glass transition temperature. For high fragilities, like $m = 170$, the divergence temperature of relaxation times, $T_{VF}$, comes close to $T_g$, while for strong glass formers, like $m = 20$, it is shifted to much lower temperatures. Assuming $\tau(T_g) = 100$ s and $\tau_0 = 10^{-14}$ s [83], the fragility dependence of the ratio $T_g/T_{VF}$ can be estimated from Eq. (1) using

$$T_g/T_{VF} = m/(m-16). \qquad (5)$$

In strong glasses with $m = 16$ (corresponding to plain Arrhenius behavior [83]), this ratio diverges with $T_{VF}$ approaching 0 K, while for large fragilities, it finally approaches unity leading to $T_g \approx T_{VF}$. The dependence of $T_g/T_{VF}$ on $m$, calculated using Eq. (5), is shown by the line in Fig. 1(b). For realistic fragility values between 20 and 170 [66], the ratio decreases almost by a factor of five on increasing $m$. In Fig. 1(b), we compare the predictions of Eq. (5) with experimental results, where the involved quantities were deduced from fits of $\tau(T)$ or $\eta(T)$ (giving $T_{VF}$, $m$, and $T_g$), from the corresponding Angell plots ($m$), and/or from heat-capacity measurements ($T_g$). Specifically, in Fig. 1(b) we added results from molecular [28,85,86,87] and ionic glass formers [88,89], polymers [49,90,91,92,93,94,95,96,97], metallic glass formers [98,99], network silicates [87,100,101,102] and other network systems [87,103,104,105,106]. Notably, some of the values on ion conductors were derived from dielectric modulus or conductivity spectra. Figure 1(b) documents that the experimental $T_g/T_{VF}$ ratios and the calculated fragility dependence using Eq. (5) are in reasonable agreement. We conclude that in fragile glass formers the kinetic glass-formation temperature approaches the critical temperature, while in strong glass formers these two characteristic temperatures are widely separated.

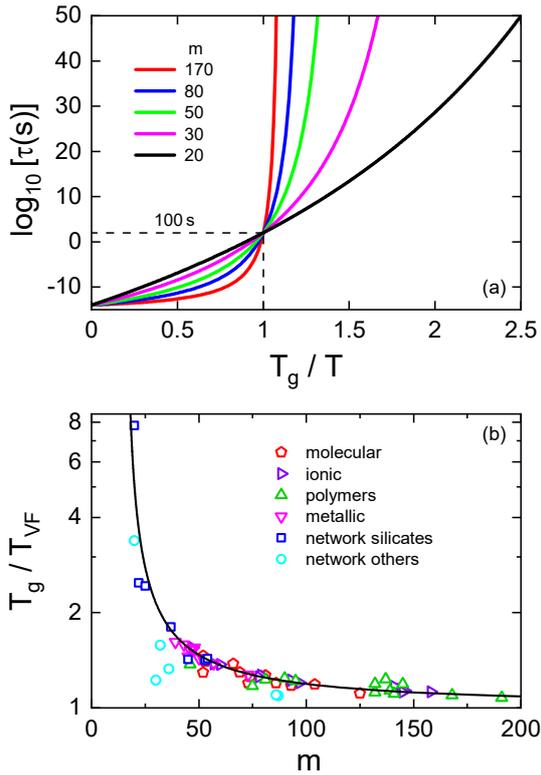

FIG 1. (a) $T_g$-normalized Arrhenius representation (Angell plot [84]) of the temperature-dependent relaxation times of supercooled liquids for fragilities $m$ between 20 and 170, calculated assuming VFT behavior, Eq. (1). (b) Fragility dependence of the ratio of glass-transition and Vogel-Fulcher temperature calculated from Eq. (5) (line). The symbols show experimental results derived for a variety of materials belonging to different classes of glass formers (see text for details).

In the following, we propose that the degree of separation of the temperature of glassy freezing from that of the extrapolated criticality as shown in Fig. 1(b) plays a fundamental role for the deviation of the PDR from unity. In Table SI [37] we have collected relevant data for a variety of glass-forming materials, for which thermal expansion, heat capacity, and compressibility were measured above and below the glass transition, and where, in addition, the specific volume at $T_g$ and the fragility were reported. We have critically reviewed existing tables [70,73,76,80,81,107,108], however, we always tried to refer to the original literature, and, in case of multiple experiments, to select the best documented results. These data allow calculating the PDR [Eq. (4)] and the two quantities $A$ and $B$ [Eqs. (2) and (3)] for various classes of amorphous materials.

Figure 2(a) shows the experimental findings from Table SI for the PDR $R$ as function of fragility $m$. Even though there is considerable scatter, it reveals a continuous increase of $R$ on decreasing fragility, spanning $R$ values from close to 1 at high, to ~ 20 at low fragilities. The scatter of $R$ could be expected in light of the variety of applied techniques and considering the experimental difficulties to determine the incremental values of the thermodynamic and mechanical quantities in the liquid and solid phases crossing glassy freezing. Notably, we provide a PDR value for amorphous silica, where no thermal expansion coefficient is reported for the supercooled liquid, by assuming the found universality $\alpha_l = 3\ \alpha_g$ where $\alpha_l$ and $\alpha_g$ are the thermal expansion coefficients in the liquid and glass state, respectively [109]. The line in Fig. 2(a) is a fit by the empirical formula $R = a/(m-16) + 1$, with a single fit parameter $a \approx 53$, supposing a divergence of $R$ at $m = 16$ and the approach of an "ideal" PDR $\approx 1$ for high fragilities. Here we assumed that the $m$ dependence of $R$ arises from the fragility dependence of the ratio $T_g/T_{VF}$, which is diverging at $m = 16$ [Fig. 1(b)].

It should be possible to trace back the found universal correlation of $R(m)$ and the deviations of $R$ from unity to the two quantities $A$ and $B$ defined in the Ehrenfest relations, Eqs. (2) and (3). As pointed out above, the first equation is valid if the configurational entropy, and the second if the volume determines $T_g$ [71]. In an ideal 2$^{nd}$ order phase transition $A = B$ must hold, resulting in $R = 1$. Taking into account the available experimental pressure dependence of the glass transition, Eq. (2) holds for most glasses, while $B$ defined in Eq. (3) is always too low [73]. In Fig. 2(b), we show the fragility dependences of $A$ and $B$. We find a gradual increase of $A$ on decreasing $m$ (circles) and, except for the lowest $m$ values, a similar gradual increase of $B$ (stars). Notably, $B$ is always smaller than $A$, leading to $R > 1$ in accord with Fig. 2(a). However, for the lowest fragilities, $m < 35$, Fig. 2(b) reveals a strong decrease of $B$, resulting in strongly increasing PDRs in Fig. 2(a). Contrary, for large $m$, $A$ and $B$ approach each other, suggesting $R \approx 1$ as expected for 2$^{nd}$ order phase transitions.

For a better insight into the fragility dependence of $R$, it is important to consider the $m$ dependences of all quantities entering the PDR [Eq. (4)], which are documented in Figs. 2(c - f). We do not plot $V_g(m)$, which is rather constant with values mostly ranging from ~0.5 – 1×10$^{-3}$ m$^3$/kg (see Table SI [37]). On decreasing $m$, the increment of thermal expansion $\Delta\alpha_p$ (c) is constant, while $\Delta c_p$ (e) reveals a small increase, but both exhibit significant downturns of approximately two orders of magnitude for the lowest fragilities. The increment of isothermal compressibility $\Delta\kappa_T$ [Fig. 2(d)] appears rather independent of $m$ when considering the data scattering. Notably, the incremental heat capacity $\Delta c_p = c_l - c_g$ (where $c_l$ and $c_g$ are the specific heat in the liquid and glass, respectively) behaves significantly different compared to the fragility dependence of the ratio $c_l/c_g$, which increases on increasing fragility [3,47]. Finally, $T_g(m)$ is presented in Fig. 2(f). On decreasing fragility, it shows a small continuous decrease, followed by a strong increase at lowest $m$ values. These detailed $m$ dependences cause the behavior of $A(m)$ and $B(m)$ and, consequently, the fragility dependence of $R(m)$. Concerning the quantity $A$ in the first Ehrenfest relation [Eq. (2)], the fragility dependences of $\Delta\alpha_p$, $\Delta c_p$, and $T_g$ almost compensate, yielding a weak, continuous evolution as function of fragility. $B$ in the second Ehrenfest relation [Eq. (3)] essentially mirrors the fragility dependence of $\Delta\alpha_p$, because $\Delta\kappa_T$ remains constant within experimental uncertainty. The low values of the incremental changes of heat capacity and thermal expansivity mostly occur in network-forming oxide glasses (see Tab. SI [37]). It should be noted that there are ideas [110,111], that such strong inorganic networks may undergo an order-disorder type liquid-liquid phase transition at much higher temperatures, thereby reducing configurational entropy considerably.

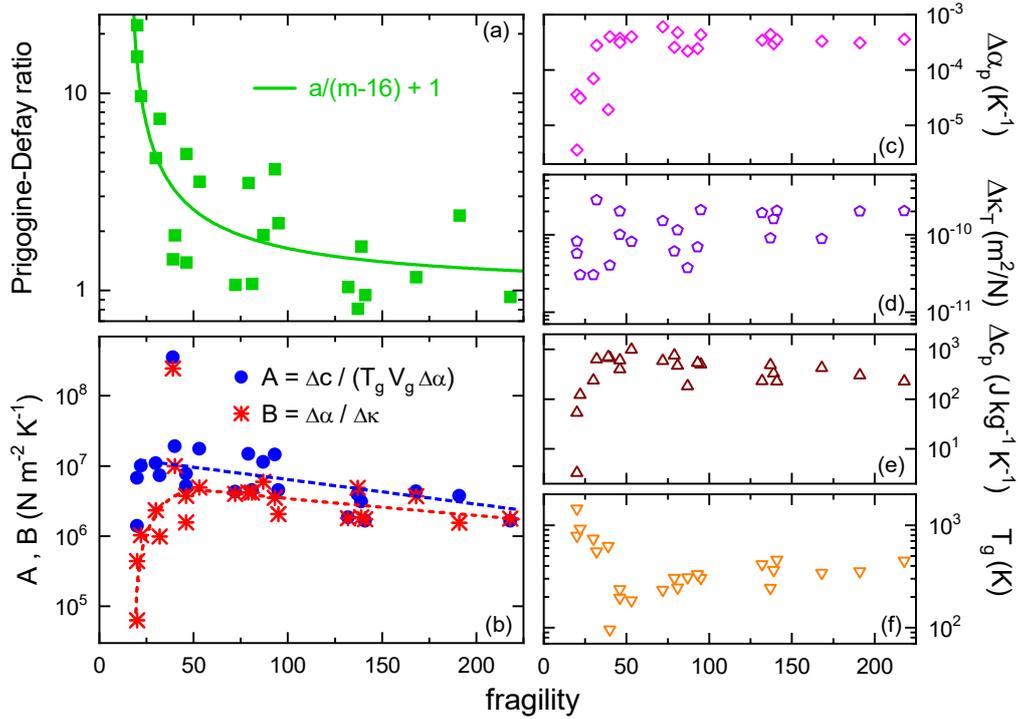

FIG 2. (a) Prigogine-Defay ratio $R$ as function of fragility $m$ for various glass-formers as listed in Table SI [37]. The line is a fit by the indicated formula, leading to $a \approx 53$. (b) Temperature dependence of $A$ and $B$, the right-hand quantities in the Ehrenfest equations [Eqs. (2) and (3)]. The dashed lines are drawn to guide the eye. (c - e) Fragility dependences of the incremental variations of thermal expansion $\Delta\alpha_p$, compressibility $\Delta\kappa_T$, and heat capacity $\Delta c_p$ at the glass transition. In (d), we excluded the value of the metallic glass, which is three orders of magnitude lower. (f) $m$ dependence of the glass-transition temperature $T_g$. All numerical data for this figure can be found in Table SI [37].

In conclusion, we found a striking correlation of $R$ and $m$, revealed in Fig. 2(a). The canonical explanation for $R > 1$ assumes that the number of order parameters involved in the glass transition is larger than one [69,70,71,72,76]. However, the observed scaling of the PDR with fragility, is in favor of an alternative interpretation: The values of $R$ larger than unity and their $m$ dependence are governed by the fragility-dependent degree of separation of glassy freezing from a hidden phase transition into an ideal low-temperature glass. For fragile glass formers, $T_g$ comes close to the ideal glass-transition temperature, approximated by $T_{VF}$ [Fig. 1(b)], and $R$ approaches unity at high $m$, as predicted by the Ehrenfest relations for 2$^{nd}$ order phase transitions. In this case, the measurements of the thermodynamic and mechanical incremental quantities are performed close to the transition into the equilibrium glass, where $R = 1$ is valid. In contrast, for strong systems with $R \gg 1$, glassy freezing appears far above this hidden phase transition and all quantities in Eq. (4) are deduced at an essentially dynamical transition far from any criticality.

In addition, we observe several further notable results: The Ehrenfest relation, Eq. (2), which results from the continuity of entropy, behaves regular and the factor $A$ reveals a minor continuous increase on decreasing fragilities [Fig. 2(b)]. This points to the importance of configurational entropy for all glass-forming liquids, independent of bonding and specific material properties. Equation (3), assuming the invariance of volume, determines $B$, which always is smaller than $A$, in accord with experimental results in literature. However, for large fragilities Fig. 2(b) documents the tendency $A \approx B$, resulting in $R = 1$, expected for ideal 2$^{nd}$ order phase transitions. On the contrary, for the lowest fragilities, $B$ exhibits a significant decrease, mainly observed for network-forming glasses (see Table SI [37]) and this behavior is responsible for $R \gg 1$. Finally, the documented $R(m)$ correlation nicely confirms that glassy freezing is driven by a hidden, underlying phase transition.

* Corresponding author. alois.loidl@physik.uni-augsburg.de

# Supplemental Material

for

# Prigogine-Defay ratio of glassy freezing scales with liquid fragility

A. Loidl,[1] P. Lunkenheimer,[1] and K. Samwer[2]

[1] Experimental Physics V, University of Augsburg, 86135 Augsburg, Germany

[2] I. Physikalisches Institut, University of Göttingen, Göttingen, Germany

**Vogel-Fulcher-Tammann (VFT) law:**

The VFT law, considered in the main text [Eq. (1)], is commonly used to describe the super-Arrhenius behavior of mean relaxation times or of viscosity in a broad range of temperatures and frequencies. It exhibits an extrapolated critical temperature below glassy freezing, where the relaxation times would diverge. As mentioned in the main text, it has been documented experimentally that this Vogel-Fulcher temperature $T_{VF}$ is very close to the Kauzmann temperature [1,2,3] where the extrapolated entropy of the supercooled liquid falls below the entropy of the fully ordered crystal, which according to popular wisdom could indicate a phase transition into an ideal low-temperature glass phase. However, there seem to exist glass-forming materials, specifically strong liquids, where this strict correlation $T_K \approx T_{VF}$ is violated [4]. Notably, the VFT law is not the only possibility to parameterize the super-Arrhenius behavior of relaxation times. For example, there are proposals by Mauro *et al*. [5] or Krausser *et al*. [6], which work equally well [7,8] with a similar number of parameters, however, without introducing a finite critical temperature. In the Krausser-Samwer-Zaccone model [6], a universal correlation between the repulsive steepness parameter $\lambda$ of the interparticle potential and the liquid fragility $m$ was found, which finally has been proven utilizing broadband dielectric spectroscopy [8].

We are aware that critical concerns regarding the use of one single VFT law, or simple extrapolations to low temperatures, as documented in Fig. 1(a) of the main text, exist. Especially, we refer to Stickel *et al*. [9,10] reporting anomalies in the variation of the relaxation times at characteristic temperatures. However, with reference to the extensive broadband dielectric spectroscopy work, including very-low frequency and ageing data, as published by Lunkenheimer, Loidl, and coworkers [11,12,13,14], it seems fair to say that the VFT law, according to Occam's razor, is the most reasonable ansatz to describe broadband relaxation data and finally is a widely accepted ansatz in the glass community.

**Glass-transition phenomena:**

As mentioned in the main text, the glass transition is a kinetic phenomenon, indicating the cross-over temperature when the system falls out of thermodynamic equilibrium and, hence, depends on the cooling rate. Figure S1 schematically shows the main characteristics of glassy freezing according to textbook knowledge, using the temperature dependence of the volume as an instructive example. A crystalline material melts in a first-order phase transition at the melting temperature $T_m$ with the appearance of considerable latent heat. If crystallization of a liquid can be avoided upon cooling, first a supercooled liquid forms, which finally, at a temperature $T_g \approx 2/3\ T_m$, undergoes glassy freezing. However, as outlined before, the glass transition is a kinetic phenomenon and depends on the cooling rate $q$, yielding lower transition temperatures for slower cooling. In the solid phases, glassy or crystalline, thermal expansion is governed by vibrational contributions only, and is similar for both solid modifications. In the supercooled-liquid phase, configurational contributions enhance the temperature dependence of the volume. On further cooling below $T_g$, assuming a constant thermal expansion, the glass finally would become denser than the crystal, roughly defining a critical temperature. Considering the excess entropy derived from specific-heat measurements, such a critical temperature was calculated by Kauzmann [15].

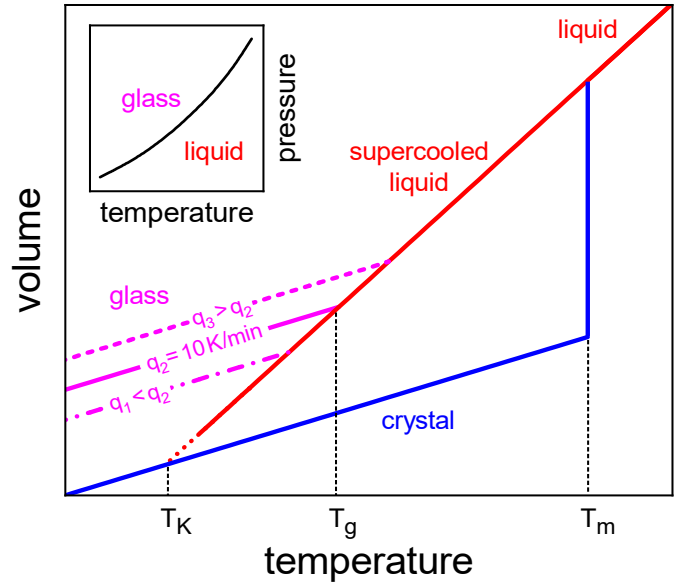

FIG. S1. Red line: Schematic temperature dependence of the volume $V$ in the liquid and supercooled-liquid states above the glass-transition. The behavior in the glass state is indicated by the magenta lines for three different cooling rates $q_1 < q_2 < q_3$. The canonically defined $T_g$ marks the glass transition for a cooling rate $q_2 \approx 10$ K/min. Blue line: $V(T)$ in the crystalline state with melting temperature $T_m$. $T_K$ indicates the Kauzmann temperature, which, in this plot, can be estimated from the extrapolated crossing of the supercooled-liquid and crystal $V(T)$ traces. Inset: Schematic second-order $(p,T)$-phase diagram between a liquid and an ideal glass, ignoring kinetic phenomena. At the phase boundary, volume $V$ and entropy $S$ of the two phases are identical. The tentative thermodynamic phase transition is hidden by kinetic phenomena and is expected at lower temperatures, close to $T_K$.

The inset of Fig. S1 shows a schematic ($p,T$) phase diagram (with $p$ being the external pressure) for a 2nd order phase-transition scenario, the line indicating a continuous transition from the liquid into the ideal glass. The first derivatives of free energy, entropy $S$ and volume $V$, are continuous across the transition, i.e., they are identical just below and above the phase boundary, while the second derivatives, specific heat $c_p$, thermal expansion $\alpha_p$, and isothermal compressibility $\kappa_T$ are discontinuous. In this schematic phase diagram, all kinetic phenomena are excluded, but could be visualized by a broadening or smearing out of the phase boundary.

**Review of published data in glass-forming liquids relevant for the Prigogine-Defay ratio:**

In Table SI we provide the relevant data for a variety of glass-forming materials, namely for systems where heat capacity, thermal expansion, and compressibility were measured above and below the glass transition and, in addition, fragilities have been determined. In addition, glass-transition temperature $T_g$ and the volume at the transition $V_g$ are indicated. From these quantities the Prigogine-Defay ratio (PDR) can be directly calculated (last column). All quantities are given in SI units and references are included directly in the table. As mentioned in the main text, we always try to refer to the original literature, and in case of multiple experiments adapted the best documented results. Partly we refer to a recent work by Lunkenheimer *et al*. [16] presenting a critical survey of existing literature data on glass-transition temperatures, fragilities, and thermal expansivities of various supercooled liquids and glasses.

Table SI documents that most of the listed polymers, polyvinylchlorid (PVC), polystyrene (PS), polypropylene (PP), polycarbonate (PC), polyarylate (PA), phenoxy (PH), polysulfone (PSF), and polyvinylacetat (PVA), have relatively large $m$ values as generally found in polymers. Exceptions are polyisobutylene (PIB) and polyethylene (PE) with significantly lower fragilities. Table SI also includes the van-der-Waals bonded, small organic-molecule system o-terphenyl (OTP) and results for OTP-OPP (o-terphenyl with o-phenylphenol) mixtures. In addition, it shows data for various hydrogen-bonded liquids, like n-propanol and glycerol, as well as for several inorganic covalent network-forming liquids, like amorphous selenium (Se), and some oxides, like $B_2O_3$, $GeO_2$, the silicate glass $26Na_2O$-$74SiO_2$, albite ($NaAlSi_3O_8$), and amorphous silica. The latter all reveal rather low fragilities, with $SiO_2$ and $GeO_2$ belonging to the strongest supercooled liquids known. So far, no reliable PDR was published for amorphous silica, mainly because the jumps in heat capacity and thermal expansion are not well documented. By analyzing the heat-capacity jump at $T_g$ from Ref. [37] and assuming that the thermal expansion in the supercooled liquid is three times that of the glass [16], we arrive at $R \approx 22.1$, which seems reliable when compared to values reported for the other oxides. We also added the ionic melt $2Ca(NO_3)_2$:$3KNO_3$ (CKN), which is an ionic mixed system close to the eutectic point and the metallic glass $Zr_{46.75}Ti_{8.25}Cu_{7.5}Ni_{10}Be_{27.5}$ (ZrTiCuNiBe). In Ref. [31] erroneously a Zr percentage of 46.25 instead of 46.75 was specified.

In addition, the values for the metallic glass, given in Tab. SI, mostly refer to the similar compound $Zr_{41.2}Ti_{13.8}Cu_{12.5}Ni_{10}Be_{22.5}$ [31]. Our slightly different $R$ value, compared to Ref. [31], results from the fact that we used the calorimetric glass-transition temperature. Already a rough inspection of Table SI reveals that the polymers mainly exhibit high fragilities and Prigogine-Defay ratios between 1 and 2 while the glass formers with low fragility tend to have higher R values.

Table SI. Glass-transition temperature $T_g$, fragility $m$, and volume $V_g$ at $T_g$. $\Delta\alpha_p$, $\Delta c_p$, and $\Delta\kappa_T$ denote incremental variations of thermal volume expansion, heat capacity, and isothermal compressibility at the glass-transition temperature. All quantities are given in SI units. The resulting PDR $R$ is presented in the last column. The relevant references are given in the table.

|  | $T_g$ (K) | $m$ | $10^3 V_g$ (m$^3$/kg) | $10^4 \Delta\alpha_p$ (K$^{-1}$) | $10^{-2} \Delta c_p$ J/(kg K) | $10^{12} \Delta\kappa_T$ m$^2$/N | $R$ |
|---|---|---|---|---|---|---|---|
| PVC | 355 [16] | 191[i] [16] | 0.73 [17] | 3.1 [16] | 3.0 [17] | 200 [17] | 2.4 |
| PS | 365 [16] | 139 [16] | 0.968 [18] | 2.97 [16] | 3.27 [18] | 160 [18] | 1.7 |
| PP | 244 [17] | 137 [19] | 1.13 [17] | 4.4 [17] | 4.8 [17] | 90 [17] | 0.81 |
| PC | 415 [16] | 132 [16] | 0.863 [20] | 3.44 [16] | 2.3 [20] | 192 [20] | 1.05 |
| PA | 450 [20] | 218[i] [21] | 0.8544 [20] | 3.59 [20] | 2.28 [20] | 202 [20] | 0.93 |
| PH | 341.2 [20] | 168 [21] | 0.8621 [20] | 3.30 [20] | 4.23 [20] | 88.7 [20] | 1.17 |
| PSF | 459.0 [20] | 141 [19] | 0.8374 [20] | 3.55 [20] | 2.26 [20] | 204 [20] | 0.95 |
| PVA | 304 [16] | 95 [16] | 0.843 [22] | 4.3 [16] | 5.0 [22] | 208.5 [22] | 2.2 |
| PIB | 195 [16] | 46 [16] | 1.05 [17] | 3.75 [23] | 4.0 [17] | 100 [17] | 1.4 |
| PE | 237 [19] | 46 [19] | 1.04 [17] | 3.14 [23] | 6.0 [17] | 200 [17] | 4.9 |
| OTP | 245 [16] | 81 [16] | 0.894 [24] | 4.76 [16] | 4.71 [24] | 114 [24] | 1.1 |
| OTP-OPP | 234 [16] | 72 [25] | 0.969 [18] | 6.0 [16] | 5.84 [18] | 150 [18] | 1.1 |
| n-propanol | 96 [16] | 35 [26] | 0.915 [27] | 4.0 [27] | 6.7 [27] | 40 [27] | 1.9 |
| Glucose | 305 [16] | 79 [16] | 0.649 [27] | 2.57 [16] | 7.53 [27] | 61 [27] | 3.5 |
| Glycerol | 185 [16] | 53 [16] | 0.752 [28] | 4.0 [16] | 9.8 [29] | 81 [30] | 3.6 |
| CKN | 333 [16] | 93 [16] | 0.456 [22] | 2.44 [16] | 5.4 [22] | 69 [22] | 4.1 |
| ZrTiCuNiBe4 | 625 [16] | 39 [16] | 0.164 [31] | 0.193 [16] | 7.03 [31] | 0.0785 [31] | 1.4 |
| Se | 310 [16] | 87 [16] | 0.2339 [32] | 2.2 [16] | 1.82 [33] | 37 [32] | 1.9 |
| $B_2O_3$ | 554 [16] | 32 [16] | 0.558 [22] | 2.77 [16] | 6.3 [22] | 280 [22] | 7.4 |
| $GeO_2$ | 787 [16] | 20 [16] | 0.278 [34] | 0.36 [16] | 0.53 [34] | 81.9 [34] | 15 |
| $26Na_2O-74SiO_2$ | 735 [35] | 30 [19] | 0.418 [35] | 0.7 [35] | 2.36 [35] | 30 [35] | 4.7 |
| Albite | 922 [16] | 22 [16] | 0.426[ii] [36] | 0.31 [16] | 1.22 [34] | 30[iii] [34] | 9.7 |
| $SiO_2$ | 1446 [16] | 20 [16] | 0.454 [34] | 0.036[iv] [16] | 0.033 [37] | 56.9 [34] | 22.1 |

[i] An upper limit of liquid fragility values was provided by Böhmer *et al.* [38] with $m < 200$. Probably more realistic maximum values for non-polymeric supercooled liquids were provided by Wang and Mauro [39] with $m < 175$ and by Wang *et al.* [40] with $m < 170$. According to these references, we believe that the fragilities cited in this table ranging from values of 218 to 20, certainly are realistic.

[ii] Volume calculated from Ref. [36], assuming the values at $T_g$ of albite synthesized with technical grade chemicals.

[iii] This value taken from Ref. [34] reveals large uncertainty. Here we took the mean value.

[iv] No reliable values for the thermal expansivity in the supercooled liquid are available in literature. We took a value of $\alpha_l = 3\, \alpha_g$ following a rule of thumb as given in Ref. [16].